\documentclass[sigconf]{acmart}
\AtBeginDocument{%
  }

\setcopyright{acmlicensed}
\copyrightyear{2018}
\acmYear{2018}
\acmDOI{XXXXXXX.XXXXXXX}
\acmConference[Conference acronym 'XX]{Make sure to enter the correct
  conference title from your rights confirmation email}{June 03--05,
  2018}{Woodstock, NY}
\acmISBN{978-1-4503-XXXX-X/2018/06}

\usepackage{graphicx}
\usepackage{amsmath}
\usepackage{url}
\usepackage{xcolor}
\usepackage{hyperref}
\usepackage{booktabs}
\usepackage{array}
\usepackage{geometry}
\usepackage[none]{hyphenat}
\usepackage{tikz}
\usetikzlibrary{positioning,fit}
\usetikzlibrary{calc}
\usetikzlibrary{arrows}  
\tikzset{every node/.style={align=center}}
\tikzset{
  box/.style={draw, rounded corners, align=center, inner sep=4pt},
  layer/.style={draw, rounded corners, fill=gray!10, inner sep=6pt, align=center},
  tee/.style={draw, rounded corners, fill=blue!6, inner sep=3pt, minimum width=2.6cm, align=center}
}
\usepackage{array}
\usepackage{booktabs}

\begin{document}


\title{Proof of Trusted Execution: \\A Consensus Paradigm for Deterministic Blockchain Finality}

\author{Kyle Habib}
\email{kyle@trillion.xyz}
\affiliation{%
  \institution{Trillion.xyz}
  \city{London}
  \country{United Kingdom}
}

\author{Vladislav Kapitsyn}
\email{vladislav@trillion.xyz}
\affiliation{%
  \institution{Trillion.xyz}
  \city{London}
  \country{United Kingdom}
}

\author{Giovanni Mazzeo}
\email{giovanni.mazzeo@uniparthenope.it}
\affiliation{%
  \institution{University of Naples 'Parthenope'}
  \city{Naples}
  \country{Italy}
}

\author{Faisal Mehrban}
\email{faisal@trillion.xyz}
\affiliation{%
  \institution{Trillion.xyz}
  \city{London}
  \country{United Kingdom}
}

\renewcommand{\shortauthors}{Trovato et al.}



\begin{abstract}
Current blockchain consensus protocols --- notably, \textit{Proof of Work} (PoW) and \textit{Proof of Stake} (PoS) --- deliver global agreement but exhibit structural constraints. PoW anchors security in heavy computation, inflating energy use and imposing high confirmation latency. PoS improves efficiency but introduces stake concentration, long-range and ``nothing-at-stake'' vulnerabilities, and a hard performance ceiling shaped by slot times and multi-round committee voting. \\
In this paper, we propose \textit{Proof of Trusted Execution} (PoTE), a consensus paradigm where agreement emerges from \emph{verifiable execution} rather than replicated re-execution. Validators operate inside heterogeneous VM-based TEEs, each running the same canonical program whose measurement is publicly recorded, and each producing vendor-backed attestations that bind the enclave code hash to the block contents. Because the execution is deterministic and the proposer is uniquely derived from public randomness, PoTE avoids forks, eliminates slot-time bottlenecks, and commits blocks in a single round of verification. We present the design of a PoTE consensus client, describe our reference implementation, and evaluate its performance against the stringent throughput requirements of the \textit{Trillion} decentralized exchange.
\end{abstract}

\keywords{Blockchain, Web3, Consensus, Trusted Execution Environments, Decentralized Finance, Verifiable Computing.}

\maketitle

\section{Introduction}

Consensus protocols have long shaped the mechanics of distributed trust, enabling independent nodes to agree on shared state without centralized control. Since Bitcoin’s release, \emph{Proof of Work} (PoW) and later \emph{Proof of Stake} (PoS) have become the dominant mechanisms securing public blockchains. Yet both carry structural limitations that surface sharply in latency-sensitive settings \cite{performance-consensus}. PoW achieves Sybil resistance through computational expenditure: miners compete to solve energy-intensive puzzles, a process that demands specialized hardware, burns staggering amounts of electricity, and introduces confirmation delays that stretch far beyond the tolerances of real-time financial systems \cite{pow}. Bitcoin’s $\approx7$\,tps throughput and hour-scale settlement expectations are emblematic of these constraints.
\\PoS improves energy efficiency by exchanging hash power for bonded stake, but it inherits a different set of bottlenecks \cite{pos}. Economic weight shapes block production, encouraging validator concentration and reinforcing the “rich-get-richer’’ dynamic. Moreover, PoS protocols remain tied to a minimum latency dictated by \emph{slot times}: a block cannot be produced or finalized faster than the cadence at which the protocol schedules committee participation. Even highly optimized PoS networks exhibit tens of seconds to minutes of confirmation delay due to multi-round attestation, fork-choice stabilization, and probabilistic finality. Long-range attacks, nothing-at-stake behavior, and complex slashing policies add further operational burden. These factors collectively make PoS a challenging foundation for high-throughput decentralized exchanges, cross-chain settlement layers, and automated financial systems that require deterministic, rollback-free finality within a sub-second envelope.
\\Layer-2 systems attempt to soften these constraints. Optimistic rollups batch execution off-chain but hinge on extended fraud windows before finality; zero-knowledge rollups compress state transitions into succinct proofs but incur heavy proving costs that delay settlement, especially for non-trivial workloads. TEE-backed rollups reduce verification overhead by executing off-chain logic inside Trusted Execution Environments with hardware-attested state \cite{teerollup}, but nearly all rely on centralized sequencers that can censor, delay, or reorder transactions. At a conceptual level, these systems still validate correctness by re-executing—or cryptographically reconstructing—the same computation across many nodes instead of verifying a single authoritative execution.
\\A different path has emerged with the broad availability of VM-based TEEs across multiple hardware vendors. Intel TDX, AMD SEV, AWS Nitro, and forthcoming ARM CCA support attestation of entire virtual machine images, enabling verifiable execution of full consensus clients rather than enclave-limited fragments. Earlier TEE-driven blockchain proposals---such as \emph{Proof of Luck} \cite{milutinovic}, \emph{Proof of Useful Work} \cite{zhang}, and SGX-centric confidential smart contract systems---assumed a single-vendor trust root. Under that model, a vulnerability in Intel SGX threatened the entire consensus mechanism. Only with the recent rise of heterogeneous, commodity VM-based TEEs has it become feasible to distribute trust across multiple attestation authorities and avoid single-vendor systemic fragility.

Motivated by this shift, we introduce \emph{Proof of Trusted Execution} (PoTE), a hardware-anchored consensus paradigm that replaces replicated re-execution with \emph{verifiable execution}. PoTE executes a canonical consensus program inside a federation of TEEs, each producing an attestation that binds a publicly recorded code measurement to the exact block contents. A block becomes final only once multiple independent vendors attest to the same transition. Because execution is deterministic and the proposer is uniquely derived from public randomness, PoTE admits no competing blocks at a given height. 
\\Our work is driven by the requirements of the \emph{Trillion}\footnote{\url{trillion.xyz}} decentralized exchange, which targets $>10k$ transactions per second (TPS). Existing PoS or rollup architectures cannot meet these latency bounds without compromising decentralization or verifiability. PoTE represents a viable alternative. To demonstrate practicality, we implement PoTE by patching the \emph{Lighthouse} Ethereum consensus client \cite{Lighthouse}. We revise the genesis configuration, replace PoS-driven fork choice with hardware-attested finality, and augment the block header with vendor-identifying metadata and attestation evidence. Our experimental evaluation across Intel TDX, AMD SEV, and an emulated ARM CCA environment shows that PoTE commits blocks in $100-150$ ms. With a $100$ ms processing window and blocks of $1,000$ transactions, PoTE reaches $\approx 10,000$ TPS, far beyond the theoretical limits of slot-based PoS systems ($\approx 83$ TPS at $12s$ block intervals). CPU and memory usage rise modestly due to enclave verification and attestation overheads. PoTE validators use slightly more CPU and about $20–40$ MB additional memory compared to native consensus.
The remainder of this work is organized as follows. Section~\ref{background} presents the background on consensus solutions and TEEs. Section \ref{relwork} overviews notable TEE-related research works in the field of blockchain. Section \ref{proofofcode} describes the design of our Proof of Trusted Execution. Section \ref{sec:implementation} provides details on the implementation of a Proof of Concept. Section \ref{usecase} presents our use case, i.e., the \emph{Trillion} Decentralized Exchange. Section \ref{evaluation} reports results from the experimental evaluation activity. Finally, Section~\ref{conclusion} concludes the document. 

\section{Background}
\label{background}

\subsection{Consensus-Based Verification}

Blockchain systems traditionally achieve correctness and trust through \emph{consensus-based replication}, ensuring all nodes agree on the same state even in the presence of adversaries. The two dominant mechanisms --- \emph{Proof of Work (PoW)} and \emph{Proof of Stake (PoS)} --- provide security through decentralization, but both face trade-offs that limit scalability, efficiency, and suitability for real-time financial applications.

\subsubsection{Proof of Work (PoW)}

PoW secures the network by requiring participants to solve computationally intensive puzzles to propose new blocks, establishing Sybil resistance through measurable energy expenditure. Bitcoin, as a canonical example, achieves roughly $\approx 7$ transactions per second with $\approx 60$ minutes to probabilistic finality, consuming over 150~TWh/year~\cite{pow}. While PoW ensures immutability and censorship resistance, it suffers from low throughput, high latency, substantial energy consumption, and hardware centralization, since mining competitiveness favors entities with specialized hardware and cheap electricity. Additionally, resources expended provide no direct utility beyond security, limiting overall efficiency.

\subsubsection{Proof of Stake (PoS)}

PoS replaces computational work with economic stake: validators lock tokens to earn the right to propose and attest blocks, and the probability of being selected scales with the amount staked. This shift dramatically reduces energy consumption and, on paper, increases throughput relative to PoW. Yet PoS systems are constrained by protocol-level timing parameters that impose hard limits on latency. For example, Ethereum’s Beacon Chain operates on fixed 12\,s slots, meaning that no block can be proposed faster than this cadence. Other networks adopt similar pacing---Cardano uses 1\,s slots but finalizes over longer epochs, while Polkadot targets 6\,s block times---all of which cap block production rates and introduce unavoidable delays before a state transition can be considered stable. Even in optimized settings, deterministic finality typically requires multiple slots, placing Ethereum’s finality in the 6--12 minute range in practice~\cite{pos-perf}. These timing constraints alone make PoS an uneasy fit for systems requiring near-instant settlement.
\\Beyond latency, PoS inherits several economic and systemic challenges. High participation thresholds---such as Ethereum’s 32\,ETH minimum---shrink the active validator set and reinforce centralization pressures. Large stakeholders accumulate proportionally larger rewards, entrenching their dominance and amplifying the “rich-get-richer’’ dynamic. Misconfigured or offline nodes risk penalties even without malicious intent, and reliance on staking services or custodial providers adds new operational and security attack surfaces. On the protocol side, validators can equivocate at negligible cost (the classic nothing-at-stake problem), complicating fork resolution. Historical key compromise enables long-range attacks, requiring checkpointing or weak-subjectivity assumptions to remain secure. Because participation depends on prior token ownership, fair bootstrapping remains difficult.

\subsection{VM-Based Trusted Execution Environments}
Trusted Execution Environments (TEEs) provide isolated execution contexts where code can run with integrity guarantees, even in the presence of a potentially compromised hypervisor or operating system. While TEEs are often associated with code and data confidentiality, in the context of PoTE, our primary concern is \emph{execution integrity} rather than secrecy. That is, we rely on the ability of TEEs to produce cryptographic attestations proving that the canonical code \(C\) is executed faithfully, rather than on preventing external observers from seeing memory contents.
\\VM-based TEEs extend traditional enclave models by protecting entire virtual machines. This allows execution of full OS stacks or complex workloads while retaining strong attestation capabilities. We focus on four representative VM-based TEE platforms in this work.
\\First, Intel Trust Domain Extensions (TDX) enable the creation of trust domains, isolated VMs whose memory and state integrity are protected. TDX provides remote attestation to prove that a VM is running authorized code. Second, AMD Secure Encrypted Virtualization (SEV) provides per-VM memory encryption but, more importantly for our purposes, attestation of VM launch parameters, allowing verifiers to confirm that a VM executes an approved image. Third, AWS Nitro Enclaves are lightweight, CPU-isolated VMs with restricted interfaces, offering attestation through Nitro’s hardware root of trust. Finally, ARM Confidential Compute Architecture (CCA), currently not publicly available, targets VM-level attestation and integrity verification across ARM-based platforms.
In our PoTE solution, the critical property provided by these TEEs is therefore \emph{attestation of code integrity}. This allows clients and verifiers to confirm that all state transitions are produced by enclaves executing the canonical code \(C\) correctly. While memory confidentiality and other protections are available in these platforms, they are not central to our threat model. The differences among the platforms primarily affect attestation interfaces, latency, and multi-vendor support, all of which influence the performance and robustness of a PoTE deployment.

\begin{table*}[h!]
\centering
\small
\renewcommand{\arraystretch}{1.3}
\scalebox{0.75}{
\begin{tabular}{
    >{\raggedright\arraybackslash}p{3.1cm}
    >{\raggedright\arraybackslash}p{4.0cm}
    >{\raggedright\arraybackslash}p{2.9cm}
    >{\raggedright\arraybackslash}p{2.8cm}
    >{\raggedright\arraybackslash}p{4.0cm}}
\toprule
\textbf{Work / System} & \textbf{TEE Usage Model} & \textbf{Consensus Type} & \textbf{Vendor Trust} & \textbf{Limitations} \\
\midrule
\textbf{Proof of Luck} \cite{milutinovic} &
TEE generates randomness for energy-efficient leader election (SGX-only) &
TEE-assisted consensus (leader election) &
Single vendor (Intel SGX) &
Probabilistic finality; vulnerable if SGX compromised; longest-chain competition remains \\

\textbf{Proof of Useful Work} \cite{zhang} &
TEE attests useful computation instead of wasteful hashing &
TEE-assisted consensus &
Single vendor (Intel SGX) &
Inherits SGX trust assumption; requires enclaves for every validator \\

\textbf{FastKitten} \cite{das} &
Confidential execution of Bitcoin transactions and wallet logic &
PoS / Bitcoin L1 & 
Single vendor (Intel SGX) &
Focus on confidentiality, not consensus security; no vendor diversity \\

\textbf{Oasis Network, Secret Network, Phala} \cite{oasisnetwork2024,secretnetwork2024,phalanetwork2024} &
Confidential smart contracts and encrypted user data inside SGX &
PoS consensus &
Single vendor (Intel SGX) &
Privacy-focused; inherit PoS limitations; dependent on SGX trust \\

\textbf{TEE Rollups} \cite{teerollup} &
Off-chain execution inside heterogeneous TEEs with attested state transitions &
L1-backed rollup finality &
Multi-vendor (SGX, SEV, TDX, TZ) &
Sequencer-based; still relies on L1 ordering; soft finality if sequencer misbehaves \\

\textbf{Our Work: PoTE} &
Deterministic execution of canonical validator code with attestation &
TEE-based consensus with deterministic finality &
Multi-vendor quorum (Nitro, SEV, TDX, CCA) &
Requires TEE availability; assumes at least one vendor remains uncompromised \\
\bottomrule
\end{tabular}}
\caption{Comparison of TEE-enabled blockchain protocols and systems. PoTE generalizes TEE-based consensus to a multi-vendor trust model with deterministic finality, eliminating probabilistic fork resolution and single-vendor dependence.}
\label{tab:related_tee_work}
\end{table*}

\section{Related Work}
\label{relwork}

The integration of TEEs into blockchain systems has attracted growing attention as a means to enhance confidentiality, performance, and verifiability. We can classify TEE-related research works according to four primary integration patterns: (i) confidential computation, (ii) TEE-assisted consensus, (iii) TEE-anchored bridges, and (iv) TEE-based oracles. 
\\Early efforts focused on embedding TEEs directly within consensus protocols. Milutinovic et al. \cite{milutinovic} proposed Proof of Luck, which employs Intel SGX enclaves to generate verifiable random numbers for energy-efficient leader election. Zhang et al. \cite{zhang} extended this notion in Proof of Useful Work, leveraging enclaves to attest that consensus participants perform meaningful computation rather than wasteful hashing. We can categorize these as TEE-based consensus protocols. It is important to note, however, that these studies were conducted during a period when Intel SGX was essentially the only widely accessible TEE platform. At that time, other TEE alternatives were unavailable. Consequently, most early designs inherited the limitations and trust assumptions of a single-vendor enclave ecosystem. In contrast, today's multi-vendor TEE landscape --- with open standards, cross-vendor attestation, and hardware diversity --- provides significantly stronger resilience and decentralization guarantees, motivating the multi-TEE \textit{PoTE} consensus explored in this work.
\\Beyond consensus acceleration, TEEs have been widely employed for confidential smart contracts and privacy-preserving computation. Das et al. propose FastKitten \cite{das} integrates enclave-protected execution into Bitcoin. Production systems such as Oasis Network \cite{oasisnetwork2024}, Secret Network \cite{secretnetwork2024}, and Phala Network \cite{phalanetwork2024} also rely on TEEs to enable confidential contract logic and user data \cite{cheng}\cite{jeanlouis2024sgxonerated}. These production-grade solutions are primarily focused on data confidentiality rather than consensus security and employ PoS mechanisms for consensus. Moreover, their trusted execution models are built exclusively on Intel SGX, representing a single-vendor trust assumption that can create systemic risks if that platform is compromised. Although these designs achieve strong privacy guarantees, they still inherit the limitations of their underlying PoS consensus and vendor-specific hardware trust.
\\A more recent line of research leverages TEEs to construct TEE rollups, in which off-chain computation and state-transition logic are executed inside enclaves and anchored to an L1 blockchain through attestation proofs. Unlike traditional zero-knowledge (ZK) or optimistic (OP) rollups, which respectively depend on cryptographic proofs or long challenge periods for fraud detection, TEE rollups rely on hardware-based attestation to verify computation correctness with minimal latency and gas cost.
The TEERollup architecture \cite{teerollup} exemplifies this approach by introducing a rollup framework that uses a committee of heterogeneous TEEs—including Intel SGX, AMD SEV, Intel TDX, and ARM TrustZone—to execute batched transactions off-chain. Each sequencer’s TEE validates and signs the resulting state transition, and the main chain accepts the first valid transition accompanied by sufficient enclave signatures. To ensure integrity, TEERollup employs on-chain attestation and registration, which eliminates the need for pairwise attestation between TEEs of different vendors, and a challenge mechanism that guarantees user fund redeemability even if sequencers or TEEs fail.

As summarized in Table \ref{tab:related_tee_work}, our approach builds upon this body of work but diverges in focus. Instead of employing a single-vendor enclave model, we propose a multi-vendor, TEE-based consensus mechanism, or PoTE, wherein enclaves from independent hardware vendors collectively form a quorum that attests to block validity. By distributing trust across heterogeneous TEEs and leveraging threshold attestation, our design mitigates single-vendor compromise while maintaining the low-energy and high-throughput advantages identified by prior TEE-based consensus studies.

\section{Proof of Trusted Execution: \\Formal Design}
\label{proofofcode}

\subsection{Overview and Objectives}
\begin{figure}[h]
	\centering
	\includegraphics [scale=0.4] {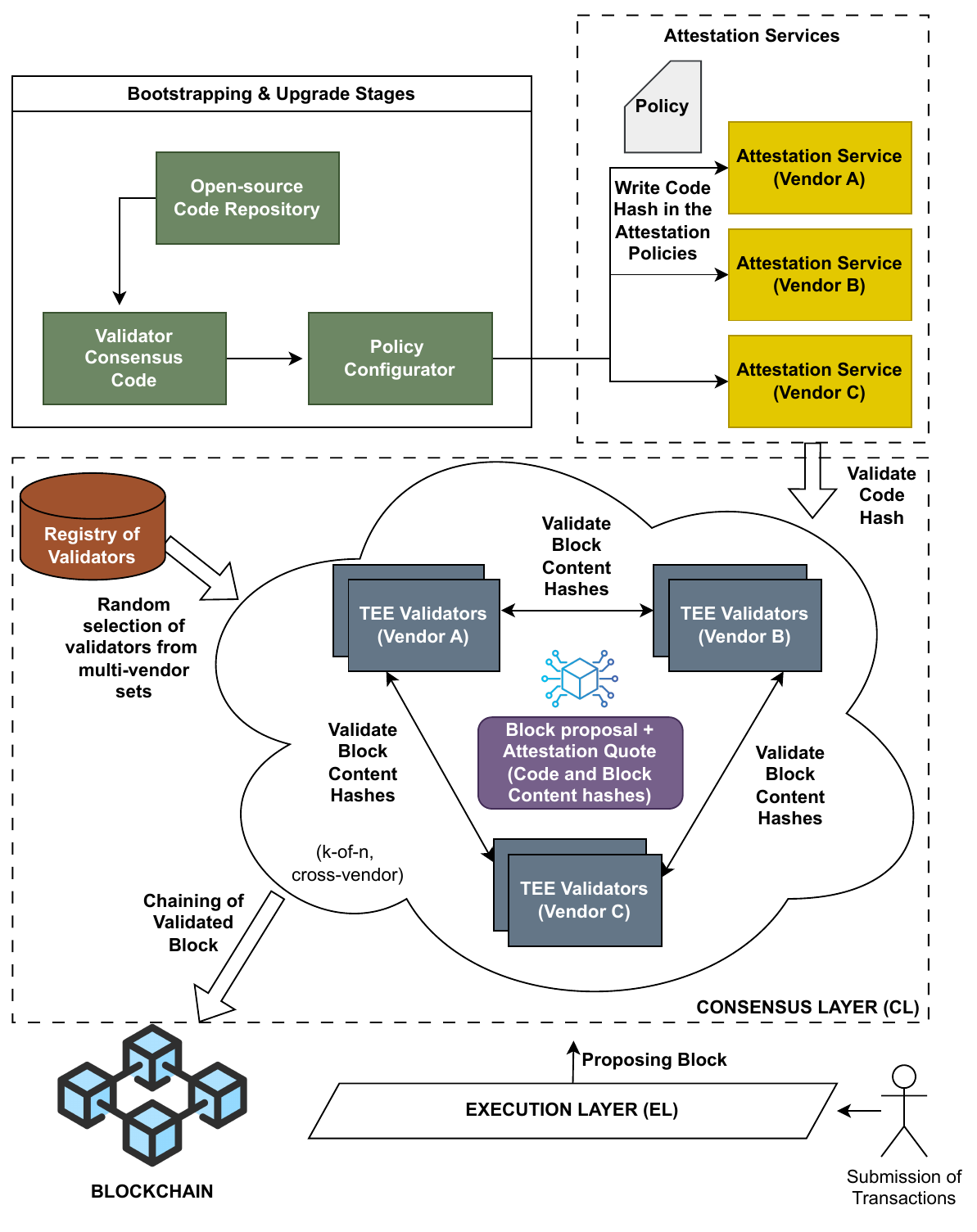}
	\caption{The PoTE Consensus Protocol}
	\label{fig:arch}
\end{figure}

Proof of Trusted Execution (PoTE) (Figure \ref{fig:arch}) is a consensus paradigm in which 
agreement arises from \emph{verifiable deterministic execution} rather 
than probabilistic fork resolution or economically weighted voting. 
Instead of re-executing each state transition across many replicas, 
validators verify a single enclave-generated transition that is 
cryptographically bound to the canonical code measurement~$\mathsf{h_C}$ 
and attested by hardware-backed roots of trust.
\\Each block is produced by a uniquely determined proposer and becomes 
final once a quorum of independent TEE vendors---formalized as a 
$k$-of-$n$ vendor threshold---attest to the correctness of the resulting 
state. At its core, PoTE assumes that at least $(n - f)$ vendors remain 
honest and that the number of compromised vendors never exceeds 
the diversity threshold $f < k$. 
This prevents any single hardware provider from unilaterally controlling 
block validation.

PoTE is built around five design objectives:

\begin{enumerate}
    \item \emph{Deterministic finality.}  
    Every block corresponds to a unique deterministic execution of the 
    canonical program~$C$. Once the required vendor attestations are 
    collected, the block becomes irrevocably final without requiring 
    fork-choice rules.

    \item \emph{Verifiable integrity.}  
    Correctness is guaranteed through hardware-rooted attestation: 
    validators confirm that an enclave executed $C$ with measurement 
    $\mathsf{h_C}$ on a specific ordered transaction batch, and that the 
    resulting block hash matches the attested commitment.

    \item \emph{Vendor-diverse trust.}  
    Safety does not depend on any single vendor. Each block requires 
    attestations from at least $k$ distinct TEE families selected from a 
    registry of $n$ authorized vendors, enabling resistance to vendor-level 
    compromise.

    \item \emph{Low-latency finalization.}  
    Finality is achieved as soon as sufficient cross-vendor attestations 
    are collected. Consensus latency is therefore dominated by enclave 
    execution and network propagation, rather than slot times or 
    committee-round protocols.

    \item \emph{Auditability and minimal re-execution.}  
    Any verifier---on-chain or off-chain---can validate a block by 
    checking its attestation set against the vendor registry and 
    $\mathsf{h_C}$, without re-executing transactions or reconstructing 
    state.
\end{enumerate}
\subsection{Threat Model and Assumptions}
\label{sec:threat-model}

PoTE rests on a hardware-anchored trust foundation provided by 
heterogeneous TEEs. Unlike traditional 
PoW- or PoS-based protocols that rely on economic assumptions, PoTE derives 
correctness from the integrity of attested execution. We assume a powerful adversary with the following capabilities:

\begin{itemize}
    \item \emph{Network-level adversary:} The adversary may delay, drop, 
    or reorder messages. Validators are assumed to operate in a partially 
    synchronous network where message delays are eventually bounded.
    
    \item \emph{Validator compromise:} The adversary may operate an 
    arbitrary number of validator nodes, may control their local operating 
    systems or hypervisors, and may attempt to tamper with enclave inputs 
    or outputs. 
    
    \item \emph{TEE-level attacks:} The adversary may exploit known 
    physical/microarchitectural attacks (e.g., TEE.fail, Foreshadow, Plundervolt), side 
    channels, or firmware vulnerabilities on specific TEE implementations. 
    These attacks may leak confidentiality or compromise local integrity of 
    an enclave instance.
    
    \item \emph{Vendor compromise:} The adversary may compromise at most 
    $f$ attestation authorities (e.g., Intel, AMD, AWS, ARM) or may forge 
    attestation evidence from those vendors. This includes key compromise 
    or a malicious vendor acting against protocol guarantees.
\end{itemize}

PoTE requires the following assumptions to hold:

\begin{itemize}
    \item \emph{TEE isolation and attestation correctness:} At least 
    $(k-f)$ distinct vendors correctly implement enclave isolation and do 
    not issue forged attestation signatures for code not matching the 
    canonical measurement~$\mathsf{h_C}$. PoTE is parameterized over 
    $(k,n)$, where $n$ is the number of registered vendors and $k$ the 
    number of vendor attestations required per block.

    \item \emph{Deterministic execution:} All enclaves execute an 
    identical, deterministic canonical program~$C$ with identical inputs. 
    Any honest enclave produces the same state transition~$\mathsf{S_{t+1}}$ 
    for the same prior state and transaction batch. Sources of 
    nondeterminism (e.g., timers, randomness, OS scheduling) are excluded 
    from the enclave environment.

    \item \emph{Bounded vendor failures:} Safety holds as long as fewer 
    than $k$ vendors are compromised. Liveness requires at least one 
    honest enclave from each of $k$ distinct vendors to be reachable.

    \item \emph{Canonical code integrity:} The code hash $\mathsf{h_C}$ 
    is published on-chain and updated only through a governed upgrade 
    process. All enclave attestations 
    must bind the produced block hash to this measurement.
\end{itemize}

PoTE does \emph{not} guarantee:

\begin{itemize}
    \item \emph{Confidentiality of enclave execution:} The threat model 
    focuses on execution integrity, not confidentiality. Side channels 
    affecting secrecy do not undermine consensus safety.

    \item \emph{Availability under cloud-level censorship:} If a cloud 
    provider disables all enclaves of a vendor, the system may enter a 
    degraded mode or require a view-change to reconfigure the vendor set.

    \item \emph{Safety under majority vendor compromise:} If $k$ or more 
    vendors collude or are compromised, they may jointly attest to false 
    state transitions. This limitation is inherent to any 
    multi-root-of-trust system.
\end{itemize}

\subsection{System Model}

The PoTE network consists of a collection of TEE-backed validator instances
(TEE-validators) denoted
\[
\mathcal{E} = \{E_{v,i}\},
\]
where each \(E_{v,i}\) is an enclave operated by some validator and instantiated on
hardware provided by vendor \(v\) (e.g., AMD~SEV, Intel~TDX, AWS Nitro,
ARM~CCA). Each enclave executes the canonical consensus program \(C\), whose
cryptographic measurement
\[
h_C = \mathsf{H}(C)
\]
is published in a registry alongside the set of authorized vendor attestation public
keys \(\{pk_v\}\). The registry defines the trust anchors that validators use to
authenticate attestation evidence.
\\Unlike Proof-of-Stake, PoTE requires no staking, bonding, or economic weighting.
A validator may join the system by launching an enclave that passes vendor-backed
attestation. During initialization, each \(E_{v,i}\) produces a signed quote:
\[
\mathsf{attest}_{v}(h_C) =
\mathsf{sig}_{v}\big(h_C \,\|\, \text{metadata}\big),
\]
certifying that the enclave runs the approved binary \(C\) on genuine hardware
recognized by vendor \(v\). Validators verify this quote \emph{off-chain} through the
vendor’s attestation service using the corresponding \(pk_v\).
\\State evolves deterministically as
\[
S_{t+1} = C(S_t, x_t),
\]
where \(S_t\) is the current state and \(x_t\) is a transaction or batch. Because
execution inside \(C\) is deterministic, all honest enclaves produce identical output
for the same input.

Each enclave signs its output:
\[
\mathsf{sig}_{E_{v,i}}\!\left(S_{t+1} \,\|\, h_C \,\|\, \mathsf{H}(B_t)\right),
\]
binding the proposed block, the canonical code hash, and the resulting state
transition.
\\Let \(\mathcal{V}\) be the set of vendors represented in the attestations attached to a
candidate block. A block is considered valid only if:
\[
|\mathcal{V}| \ge 3
\quad\text{and}\quad
\forall v \in \mathcal{V} :
\mathsf{Verify}_{pk_v}\!\big(\mathsf{attest}_{v}(h_C)\big) = 1.
\]
This requirement ensures that no block can be finalized unless enclaves from at
least three independent hardware vendors corroborate the transition, eliminating
single-vendor fragility and significantly raising the bar for adversarial compromise.

\subsection{Proposer Selection}
At each block height \(t\), PoTE designates exactly one enclave as the proposer.
This selection is deterministic, publicly verifiable, and independent of stake or
economic weight.
\\A randomness seed \(r_t\) is derived from publicly observable values, such as a VRF
beacon or a hash chain rooted in previous blocks. All validators compute \(r_t\)
independently, ensuring a common, unpredictable source of entropy that cannot be
biased by any single party.
\\To guarantee hardware diversity over time, proposer selection proceeds in two
steps. First, the seed \(r_t\) is used to uniformly sample a TEE vendor class
\(v \in \mathcal{V}_{\text{auth}}\), where \(\mathcal{V}_{\text{auth}}\) is the set of authorized vendors
(e.g., SEV, TDX, Nitro, CCA). This prevents long runs of blocks produced by a
single hardware stack and avoids any vendor dominating the chain.
\\Second, among all enclaves belonging to vendor \(v\), the seed deterministically
selects one enclave \(E_{v,i}\). Because all validators apply the same deterministic
function to the same seed, they all elect the same proposer without requiring
communication or coordination.
\\Only the designated proposer may assemble the block for height \(t\). Any block
originating from a non-elected enclave is immediately rejected. Combined with
deterministic execution inside the enclave, this guarantees that no two valid
conflicting blocks can exist at the same height.
\\Unlike Proof-of-Stake protocols, PoTE does not rely on slot times, epochs,
committees, or fork-choice rules. The proposer is uniquely and deterministically
defined for each height, and finality is determined exclusively by multi-vendor
attestation rather than probabilistic stabilization.

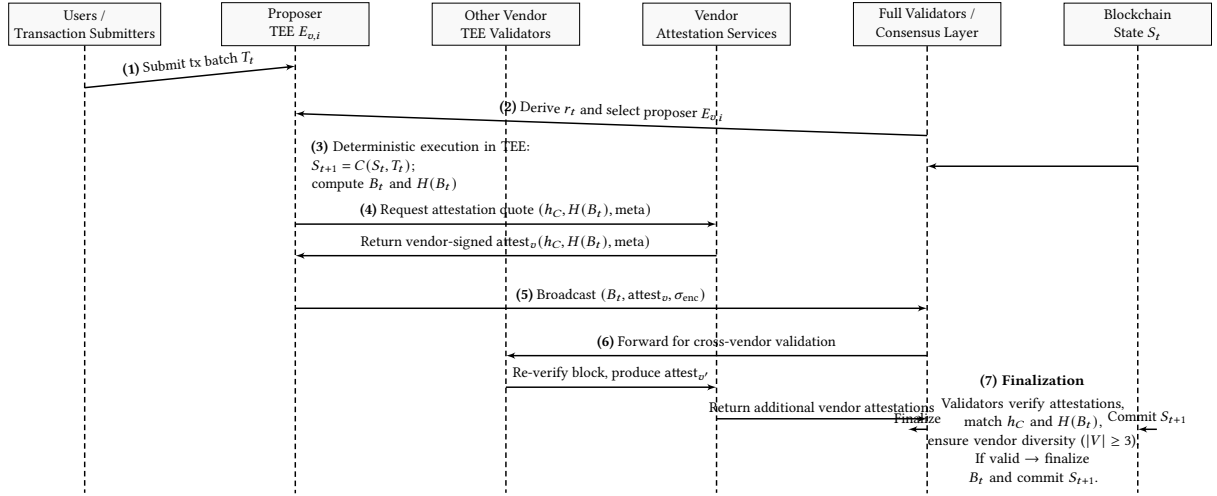
\begin{figure*}[t]
  \centering
  \scalebox{0.7}{   
  \begin{tikzpicture}[
    >=latex',
    lifeline/.style={rectangle,draw,minimum width=2.8cm,minimum height=0.8cm,fill=gray!5},
    timeline/.style={densely dashed,thick},
    call/.style={-latex',thick},
    node distance=3.0cm,
    font=\small
  ]

  \node[lifeline] (client)      at (0,0)   {Users /\\ Transaction Submitters};
  \node[lifeline] (proposer)    at (4,0)   {Proposer\\TEE $E_{v,i}$};
  \node[lifeline] (others)      at (8,0)   {Other Vendor\\TEE Validators};
  \node[lifeline] (attest)      at (12,0)  {Vendor\\Attestation Services};
  \node[lifeline] (validators)  at (16,0)  {Full Validators /\\ Consensus Layer};
  \node[lifeline] (chain)       at (20,0)  {Blockchain\\State $S_t$};

  \draw[timeline] (client.south) -- ++(0,-8.5);
  \draw[timeline] (proposer.south) -- ++(0,-8.5);
  \draw[timeline] (others.south) -- ++(0,-8.5);
  \draw[timeline] (attest.south) -- ++(0,-8.5);
  \draw[timeline] (validators.south) -- ++(0,-8.5);
  \draw[timeline] (chain.south) -- ++(0,-8.5);

  \draw[call] (client.south) ++(0,-0.8) --
    node[above,sloped]{\textbf{(1)} Submit tx batch $T_t$}
    (proposer.south |- 0,-0.8);

  \draw[call] (validators.south) ++(0,-1.7) --
    node[above,sloped]{\textbf{(2)} Derive $r_t$ and select proposer $E_{v,i}$}
    (proposer.south |- 0,-1.7);

  \node[align=left,anchor=west,text width=5.2cm] at (4.2,-2.7) {
    \textbf{(3)} Deterministic execution in TEE:\\
    $S_{t+1} = C(S_t, T_t)$;\\
    compute $B_t$ and $H(B_t)$
  };
  \draw[call] (chain.south |- 0,-2.7) -- ++(-4,0);

  \draw[call] (proposer.south |- 0,-3.8) --
    node[above,sloped]{\textbf{(4)} Request attestation quote $(h_C, H(B_t),\text{meta})$}
    (attest.south |- 0,-3.8);

  \draw[call] (attest.south |- 0,-4.4) --
    node[above,sloped]{Return vendor-signed $\mathsf{attest}_v(h_C,H(B_t),\text{meta})$}
    (proposer.south |- 0,-4.4);

  \draw[call] (proposer.south |- 0,-5.4) --
    node[above,sloped]{\textbf{(5)} Broadcast $(B_t, \mathsf{attest}_v, \sigma_{\text{enc}})$}
    (validators.south |- 0,-5.4);

  \draw[call] (validators.south |- 0,-6.3) --
    node[above,sloped]{\textbf{(6)} Forward for cross-vendor validation}
    (others.south |- 0,-6.3);

  \draw[call] (others.south |- 0,-6.9) --
    node[above,sloped]{Re-verify block, produce $\mathsf{attest}_{v'}$}
    (attest.south |- 0,-6.9);

  \draw[call] (attest.south |- 0,-7.5) --
    node[above,sloped]{Return additional vendor attestations}
    (validators.south |- 0,-7.5);

 \node[align=center, text width=4.5cm] (step7) at (18,-7.7) {
      \textbf{(7) Finalization}\\[2pt]
      Validators verify attestations, match $h_C$ and $H(B_t)$,\\
      ensure vendor diversity ($|V|\!\ge\!3$).\\
      If valid → finalize $B_t$ and commit $S_{t+1}$.
  };

  \draw[call] (validators.south |- 0,-7.7) --
      node[above,sloped]{Finalize}
      (step7);

  \draw[call] (step7) --
      node[above,sloped]{Commit $S_{t+1}$}
      (chain.south |- 0,-7.7);

  \end{tikzpicture}}

  \caption{PoTE block lifecycle. A uniquely selected proposer enclave executes the canonical program $C$ on state $S_t$ and transaction batch $T_t$, obtains a hardware attestation binding the code hash $h_C$ and block hash $H(B_t)$, and broadcasts the attested block. Validators gather cross-vendor re-attestations and finalize once a multi-vendor quorum is met, producing deterministic, fork-free state $S_{t+1}$.}
  \label{fig:pote-lifecycle}
\end{figure*}

\subsection{Protocol Flow}

PoTE achieves consensus through a single, deterministic pipeline that replaces
multi-round voting with attested execution and cross-vendor verification. The
lifecycle (Figure \ref{fig:pote-lifecycle}) of a block at height \(t\) proceeds through the following phases.

\textit{1) Proposer determination.}
All validators compute the public randomness seed \(r_t\) and derive the unique
proposer \(E_{v,i}\) for the round using the vendor-aware selection function
described in Section~IV-C. No coordination or committee formation is required.

\textit{2) Deterministic execution.}
The proposer enclave retrieves the current ledger state \(S_t\) and the ordered
transaction batch \(T_t\). It executes the canonical program \(C\) inside its TEE,
producing the next state:
\[
S_{t+1} = C(S_t, T_t).
\]
Because \(C\) is deterministic, any honest enclave would produce exactly the same
output for the same input.

\textit{3) Block assembly and attestation.}
The proposer constructs a candidate block \(B_t\) containing \(S_t\), \(T_t\),
the resulting state commitment, and metadata. The enclave then generates an
attestation quote binding:
\begin{itemize}
  \item the canonical code hash \(h_C\),
  \item the block hash \(\mathsf{H}(B_t)\),
  \item protocol metadata (height, chain identifier, nonce).
\end{itemize}
The enclave signs the full block, producing a verifiable artifact that ties the
attestation to the block contents.

\textit{4) Broadcast and distributed verification.}
The proposer broadcasts \((B_t, \text{quote}, \sigma)\) to all validators. Each validator
verifies the attestation using the vendor’s public key and attestation service,
checks the code measurement against \(h_C\), recomputes the block hash, and
verifies the enclave signature \(\sigma\).

\textit{5) Cross-vendor re-attestation.}
Validators belonging to distinct TEE vendors independently re-execute the final
verification step within their enclaves. Each produces a matching attestation that
confirms the block’s correctness. These attestations accumulate until the
multi-vendor threshold is met.

\textit{6) Finalization.}
Once attestations from at least three independent vendor families are collected,
\(B_t\) is declared final. No fork-choice rule, no probabilistic stabilization, and
no further rounds are needed. The chain advances deterministically to state
\(S_{t+1}\).

\textit{7) Chaining.}
The next round begins by deriving the seed \(r_{t+1}\) from publicly visible
state—including the hash of \(B_t\)—ensuring that proposer selection cannot be
manipulated by an adversary.

This single-round flow yields deterministic finality with latencies governed
primarily by attestation generation and network propagation, enabling PoTE to
operate at sub-second timescales without altering the execution semantics or
trust model of higher-level applications.

\subsection{Block Construction and Chaining}
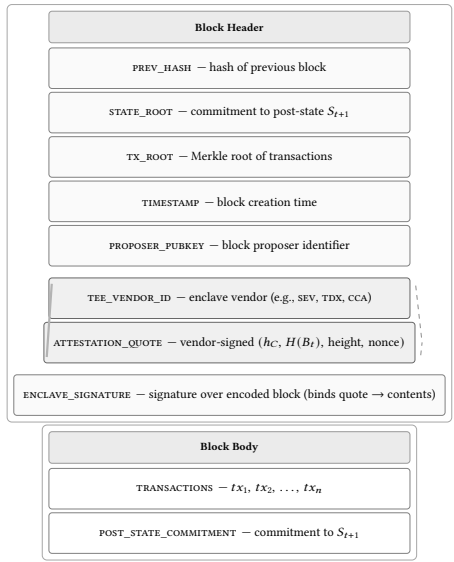
\begin{figure}[t]
\centering

\newcommand{\BlockWidth}{8.0cm} 
\scalebox{0.6}{
\begin{tikzpicture}[
  font=\small,
  row/.style = {rectangle, draw=black!50, rounded corners=2pt, fill=white,
                minimum height=0.9cm, minimum width=\BlockWidth, inner sep=6pt, align=left},
  rowdim/.style = {row, draw=black!50, fill=black!2},
  rowpote/.style = {row, draw=black!60, fill=black!6}, 
  titlebar/.style = {rectangle, rounded corners=2pt, fill=black!8, draw=black!40,
                     minimum height=0.7cm, minimum width=\BlockWidth, inner sep=5pt, font=\footnotesize\bfseries, align=left},
  slim/.style = {inner sep=0pt, outer sep=0pt},
]

\node[titlebar] (hdrtitle) {Block Header};

\node[rowdim, below=1mm of hdrtitle] (prev)  {\textsc{prev\_hash}\, — hash of previous block};
\node[rowdim, below=0.7mm of prev]   (state) {\textsc{state\_root}\, — commitment to post-state $S_{t+1}$};
\node[rowdim, below=0.7mm of state]  (tx)    {\textsc{tx\_root}\, — Merkle root of transactions};
\node[rowdim, below=0.7mm of tx]     (time)  {\textsc{timestamp}\, — block creation time};
\node[rowdim, below=0.7mm of time]   (prop)  {\textsc{proposer\_pubkey}\, — block proposer identifier};

\node[rowpote, below=2.5mm of prop]  (vendor) {\textsc{tee\_vendor\_id}\, — enclave vendor (e.g., \textsc{sev}, \textsc{tdx}, \textsc{cca})};
\node[rowpote, below=0.7mm of vendor] (quote) {\textsc{attestation\_quote}\, — vendor-signed $(h_C,\, H(B_t),\, \text{height},\, \text{nonce})$};
\draw[black!30, line width=1.4pt] ($(vendor.west)+(0.07,0.32)$) -- ($(quote.west)+(0.07,-0.32)$);

\node[rowdim, below=2.5mm of quote] (sig) {\textsc{enclave\_signature}\, — signature over encoded block (binds quote $\rightarrow$ contents)};

\node[draw=black!35, rounded corners=3pt, fit=(hdrtitle)(prev)(state)(tx)(time)(prop)(vendor)(quote)(sig),
      inner sep=4pt] (headerframe) {};

\node[titlebar, below=2mm of headerframe] (bodytitle) {Block Body};
\node[row, below=1mm of bodytitle] (txs) {\textsc{transactions}\, — $tx_1,\, tx_2,\, \dots,\, tx_n$};
\node[row, below=0.7mm of txs]    (post) {\textsc{post\_state\_commitment}\, — commitment to $S_{t+1}$};

\node[draw=black!35, rounded corners=3pt, fit=(bodytitle)(txs)(post), inner sep=4pt] (bodyframe) {};

\node[slim, anchor=west] (note) at ($(quote.east)+(0.15,0)$) {};
\draw[dashed, black!45] ($(vendor.east)+(0.12,0.28)$) -- (note.west);
\draw[dashed, black!45] ($(quote.east)+(0.12,-0.28)$) -- (note.west);

\end{tikzpicture}}

\caption{Extended block structure with \emph{PoTE}-specific header fields highlighted. The additional \texttt{tee\_vendor\_id} and \texttt{attestation\_quote} bind the canonical code hash $h_C$ and block content hash $H(B_t)$ to the proposal, enabling deterministic verification without fork-choice.}
\label{fig:pote-block-extended-pretty}
\end{figure}

In PoTE, a \emph{block} (Figure \ref{fig:pote-block-extended-pretty}) represents a deterministic state transition produced by executing the canonical code \(C\) on a set of transactions. Let \(\mathcal{T}_t = \{x_{t,1}, x_{t,2}, \dots, x_{t,n}\}\) denote the ordered set of transactions collected for block \(B_t\). Each enclave executes \(C\) on the current ledger state \(S_t\) and sequentially applies each transaction \(x_{t,i}\) to produce a new state \(S_{t+1} = C(S_t, \mathcal{T}_t)\). The resulting state, along with the transaction list and attestation evidence, forms the content of the block:
\[
B_t = \left(S_t, \mathcal{T}_t, S_{t+1}, \{\mathsf{attest}_{v}(S_{t+1} \| h_C)\}_{v \in Q}\right),
\]
where \(Q\) is a quorum of distinct vendor enclaves providing valid attestations.
Each block is linked to its predecessor through a standard cryptographic hash:
\[
\mathsf{prev\_hash}_t = \mathsf{H}(B_{t-1}),
\]
and the block header includes \(\mathsf{prev\_hash}_t\), the canonical code hash \(h_C\), a timestamp, and a digest of attestation signatures. This ensures that the chain is tamper-evident: any modification of a previous block changes the hash, invalidating all descendant blocks unless accompanied by forged attestations from a quorum of vendors.
\\Block finalization proceeds once the verifier confirms that a quorum of enclave attestations match the canonical measurement \(h_C\) and produce identical state \(S_{t+1}\). Because enclaves are deterministic, all verified replicas of the block contain the same transaction outcomes and state, eliminating forks under normal operation. Unlike PoW or PoS, no leader election or probabilistic finality is required; the chain grows linearly as each attested block is appended.

\subsection{Security Properties}

PoTE assumes a partially trusted hardware model in which TEEs correctly enforce isolation and vendor attestation keys cannot be forged without compromising the vendor’s hardware root of trust. Under these assumptions, PoTE provides five core verification guarantees: attestation correctness, state integrity, signature soundness, round freshness, and multi-vendor diversity.

\paragraph*{Attestation verification.}
Every validator receives a block proposal \((B, \text{quote}, \sigma)\) and first verifies the vendor attestation. The quote is checked against the vendor’s root public key and must show that the enclave executed the canonical code with measurement \(h_C\). The validator also checks firmware and TCB levels, and ensures that the enclave is not revoked. The quote carries attested user data from which the validator extracts:
\[
(pk_{\mathsf{block}},\, C,\, \mathsf{height},\, \mathsf{chain\_id},\, \mathsf{nonce}).
\]

\paragraph*{Commitment check.}
The quote binds the enclave to a commitment \(C\) representing the hash of the proposed block. The verifier recomputes
\[
C' = H(\mathrm{encode}(B))
\quad\text{and requires}\quad C' = C.
\]
If a malicious proposer modifies the block after attestation, the hash no longer matches, and the block is rejected immediately.

\paragraph*{Signature check.}
The block carries a signature \(\sigma\) generated with the ephemeral enclave key \(pk_{\mathsf{block}}\). The verifier checks that
\[
\mathsf{Verify}_{pk_{\mathsf{block}}}(\sigma, \mathrm{encode}(B)) = 1,
\]
which proves that the enclave producing the attestation also signed exactly this block. If an attacker attempts to keep the attestation but alter any byte of \(B\), signature verification fails.

\paragraph*{Round freshness.}
To prevent replay attacks, the attested user data includes a protocol nonce, chain identifier, and block height. Validators require
\[
(\mathsf{height},\, \mathsf{chain\_id},\, \mathsf{nonce})\]
\[\quad\text{match the expected values of the current round.}
\]
A quote from a previous block or different chain therefore cannot be reused to authorize a new proposal.

\paragraph*{Diversity rule.}
PoTE enforces multi-vendor safety by requiring that each finalized block be validated by enclaves from at least three distinct TEE families. Validators track which TEE technologies have contributed valid triples \((B, \text{quote}, \sigma)\) in the current round. A block is accepted only once the diversity requirement is satisfied, ensuring that no single vendor compromise can authorize invalid state transitions.

To summarize, if a malicious validator attempts to ``attach a correct quote but send a different block,'' at least one check fails. If the attacker modifies \(B\) after attestation, the recomputed hash \(C' = H(\mathrm{encode}(B))\) no longer equals the attested \(C\), causing a commitment failure. If the attacker keeps \(C\) but alters \(B\), then \(\sigma\) is no longer a valid signature over the modified block, causing a signature failure. If the attacker attempts to replay an old quote, the height, chain identifier, or nonce mismatch triggers a freshness failure. Thus, every accepted block must be (i) produced by a genuine enclave executing \(C\), (ii) bound to its exact contents, and (iii) attested within the correct round.

\section{Implementation Details}
\label{sec:implementation}
We implemented a proof-of-concept for PoTE by extending the Consensus Layer (CL) of a
widely used Ethereum client. Specifically, we maintain a dedicated fork of
the \texttt{Lighthouse} client written in Rust.\footnote{\url{https://github.com/trillion-xyz/PoTE-lighthouse}}
Our extensions augment the beacon block header with two new fields:
(i)~the TEE vendor identifier (1 byte), and (ii)~a raw attestation quote produced by the
enclave (maximum 8192 bytes). These fields bind each block to a specific hardware vendor and to the
canonical enclave code hash~$h_C$, enabling downstream verifiers to check both
the trust root and code integrity of the proposed block.
\\Because the Ethereum CL encodes block headers and state with
SSZ serialization, the additional header fields required end-to-end changes to
the serialization and deserialization logic. To preserve compatibility with
Lighthouse’s internal types and with the deposit contract used to initialize a
beacon chain, we introduced corresponding changes in serialization and genesis-generation components. This required maintaining PoTE-specific forks of the
\texttt{go-eth2-client} and \texttt{eth-beacon-genesis}
libraries.\footnote{\url{https://github.com/trillion-xyz/PoTE-go-eth2-client}}\footnote{\url{https://github.com/trillion-xyz/PoTE-eth-beacon-genesis}}
These modifications ensure that PoTE-compatible chains can bootstrap with the
correct block header format and SSZ encoding rules from genesis onward.
\\Beyond structural changes to the block format, we extend the block validation
pipeline within Lighthouse to enforce PoTE’s multi-vendor security model.
During block import, the client extracts the attestation quote, verifies its
signatures and freshness, and checks that the reported measurement matches
$h_C$. A block is accepted only if a quorum of valid attestations is present
and the set of attestations satisfies the required vendor diversity threshold.
This logic replaces traditional consensus voting and fork-choice rules with a
verifiable, single-round commit: once a block carries a quorum of valid
cross-vendor attestations, any PoTE verifier (including an on-chain verifier)
can deterministically accept it without re-execution.
\\To support multi-node experimentation and reproducible testnets, we rely
on \texttt{Kurtosis} to orchestrate Lighthouse instances and auxiliary services.
However, PoTE requires exposing a hardware device (e.g., \textit{/dev/tdx-guest}) directly inside the internal containers launched by Kurtosis, a
capability not supported by the upstream framework. To enable this, we forked
the \texttt{Kurtosis} repository and implemented device-sharing support for
nested containers, and subsequently opened a pull request upstream.\footnote{\url{https://github.com/trillion-xyz/PoTE-kurtosis}}
This extension allows our PoTE-enabled Lighthouse nodes to access the emulated TEE device during integration tests and distributed experiments.

\section{Use Case: The \textit{Trillion} Decentralized Exchange}
\label{usecase}
The design of PoTE was developed in direct response to the concrete performance and trust requirements of \textit{Trillion}, a decentralized exchange engineered for latency-sensitive and high-throughput markets.\footnote{\url{https://trillion.xyz/}} Trillion aims to support a broad spectrum of financial primitives—spot markets, perpetuals, options, algorithmic strategies, and real-world assets—while achieving execution properties typically associated with centralized venues: sub-second order processing, deterministic settlement, and throughput on the order of tens of thousands of transactions per second. These expectations, when transplanted into a decentralized setting, expose structural limitations in existing blockchain consensus protocols. The gap between Trillion's requirements and the capabilities of traditional designs ultimately guided the architecture of PoTE, motivating a shift toward deterministic execution, single-round finality, and hardware-anchored verification.%
\\Latency is the most immediate constraint. A matching engine designed for high-frequency and algorithmic traders cannot absorb the delays inherent in slot-based proposer schedules, fork-choice convergence, or multi-round attestation procedures. These timing components are fundamental to Proof-of-Stake protocols and impose an irreducible lag between transaction submission and confirmed state updates. For a system like Trillion, even short-lived forks or probabilistic finality windows translate directly into degraded market quality and inconsistent execution outcomes. PoTE avoids these bottlenecks by grounding consensus in deterministic enclave execution: a uniquely derived proposer produces exactly one block, and finality is reached as soon as a quorum of heterogeneous TEEs attest to the correctness of the output. In effect, consensus latency becomes a function of hardware attestation and network propagation rather than protocol-mandated waiting periods.%
\\Correctness is equally critical. Without cryptographic guarantees of execution integrity, a matching engine becomes a de facto trust anchor—an outcome incompatible with Trillion’s decentralized security model. PoTE addresses this concern by embedding integrity proofs directly into the block production pipeline. Each block is generated inside a TEE running a canonical program whose measurement~$h_C$ is publicly registered, and the block header carries an attestation binding the enclave’s identity, the code hash, and the block contents. Any validator or on-chain verifier can independently confirm that the block was produced by authorized code executing on genuine hardware, removing reliance on operator honesty or opaque off-chain mechanisms.%
\\Finally, institutional-grade settlement requires resilience to hardware compromise. Depending on a single TEE vendor would create a brittle trust root, as the failure or compromise of one attestation authority could undermine the entire system. PoTE incorporates vendor diversity into its safety criteria: a block is accepted only once enclaves from at least three independent hardware families corroborate the same state transition. This requirement prevents any single vendor—intentionally or accidentally—from exerting unilateral influence over settlement. For a globally deployed DEX such as Trillion, which must operate across heterogeneous cloud environments, this multi-vendor attestation model is essential for maintaining strong and defensible trust guarantees.%

\section{Experimental Evaluation}
\label{evaluation}
We empirically evaluate PoTE to demonstrate that verifiable and deterministic
finality can be achieved with sub-second commit latency and without consensus
re-execution. Our experiments compare PoTE against a native Proof-of-Stake
(PoS) validator cluster running unmodified Lighthouse, providing an
apples-to-apples benchmark. 

\begin{table}[t]
\centering
\caption{PoTE Azure-based testbed configuration (Ubuntu 24.04 LTS).}
\scalebox{0.85}{
\begin{tabular}{l p{2.8cm} p{2.8cm}}
\toprule
\textbf{Azure VM} & \textbf{Resources} & \textbf{TEE Backend} \\
\midrule
DCasv5 & 8\,vCPUs, 32\,GiB RAM & AMD SEV \\
DCesv6 & 8\,vCPUs, 32\,GiB RAM & Intel TDX \\
Dpdsv6 & 8\,vCPUs, 32\,GiB RAM & QEMU-emulated ARM CCA \\
\bottomrule
\end{tabular}}
\label{tab:testbed}
\end{table}

\subsection{Testbed}
We deployed a private PoTE beacon chain using our forked version of \texttt{Kurtosis}\footnote{\url{https://github.com/trillion-xyz/PoTE-kurtosis}} to provision
reproducible clusters and inject controlled network conditions. The testbed (Table \ref{tab:testbed}) consists of three Azure cloud nodes in West Europe located in the same availability zone. To model a heterogeneous multi-vendor
committee, each node has a different TEE: Intel
TDX, AMD SEV, and (emulated) Arm CCA. The emulated CCA validators follow the
same attestation and validation pipeline as hardware-backed enclaves, but their
quotes are signed using a software model of the CCA attestation authority. This
lets us exercise PoTE's vendor-diversity threshold even where cloud CCA
hardware is not yet generally available. On top of these three heterogeneous nodes, we vary the total number of validators, increasing the validator population per node while keeping the number of physical nodes constant. This setup allows us to evaluate how PoTE scales when more validators participate in attestation and verification, without changing the underlying network or TEE distribution.
\\All validators run our modified \texttt{Lighthouse} client and use the
extended block header format and SSZ codecs described in
Section~\ref{sec:implementation}. Enclaves produce block-level attestations
binding the canonical code hash~$h_C$ to the block body. During block import,
validators verify the attestation signatures and freshness and enforce PoTE's
multi-vendor quorum rule.

\subsection{Observed Metrics}
For each configuration of validators, we collect a set of performance metrics that directly reflect PoTE's consensus properties and the role of hardware attestation in block production:
\begin{itemize}
    \item \textit{Block commit latency} -- it is the time from when a block is first observed to when it becomes the canonical head block. It measures how quickly a proposed block is committed to the chain.
    \item \textit{Transactions per second} -- it represents the effective application-level throughput that the system can sustain assuming 1k transactions in each block. We compute TPS as the number of transactions contained in a committed block divided by the measured block commit latency. This metric captures how improvements in consensus latency—such as those enabled by PoTE's slotless design—translate into higher end-to-end throughput and the ability to sustain workloads such as high-volume DEX trading.
    \item \textit{Average CPU usage} -- it reflects how much processing power validators consume when handling blocks, signatures, networking, or disk operations. 
    \item \textit{Peak Memory usage} -- it captures the peak amount of RAM required by validators during their operation, e.g., to store state, buffers, block data, and cryptographic structures.
\end{itemize}

\begin{figure}[h]
	\centering
	\includegraphics [scale=0.3] {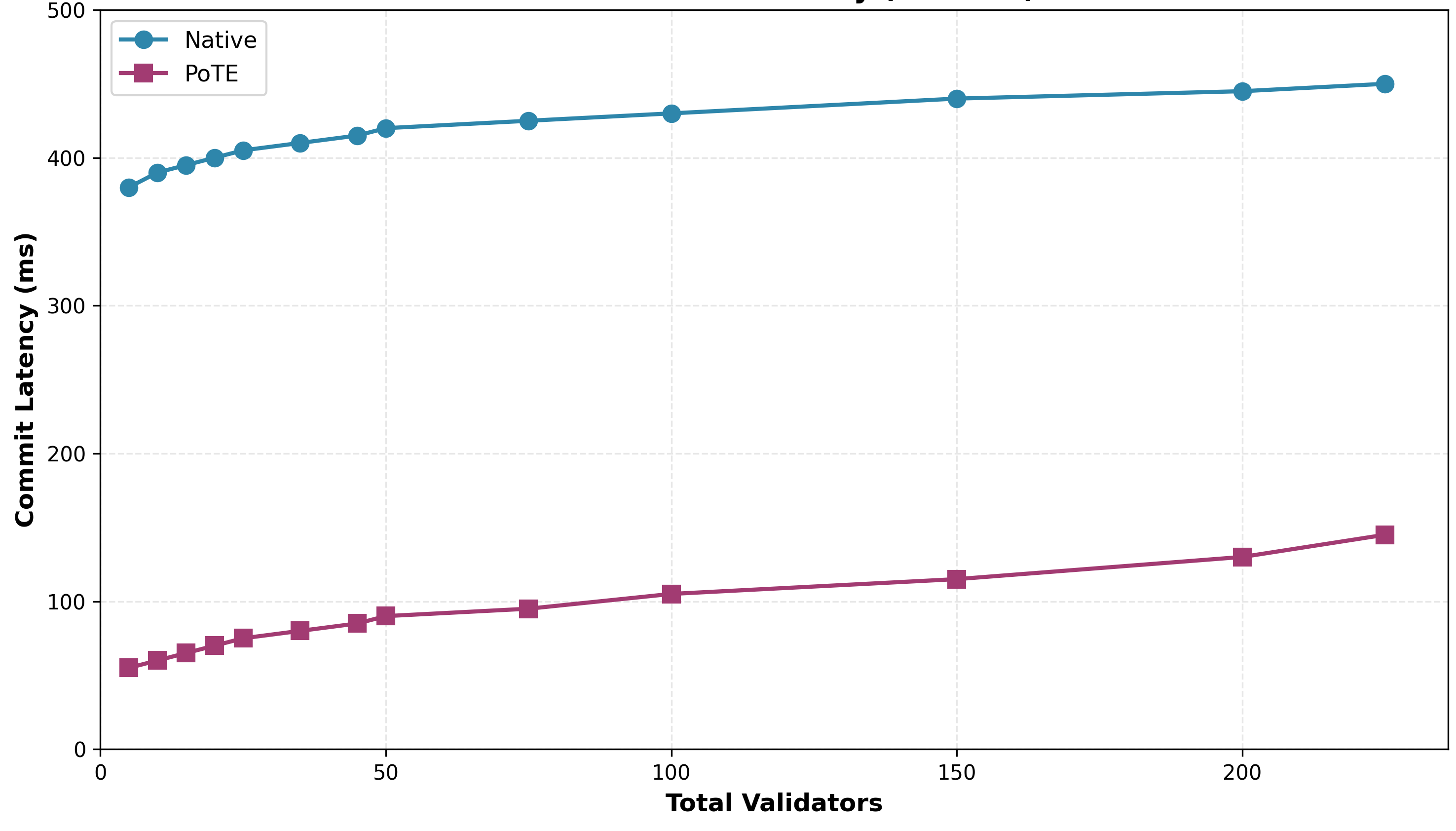}
	\caption{Block Commit Latency}
	\label{fig:commit-latency}
\end{figure} 

\begin{figure*}[h]
	\centering
	\includegraphics [scale=0.4] {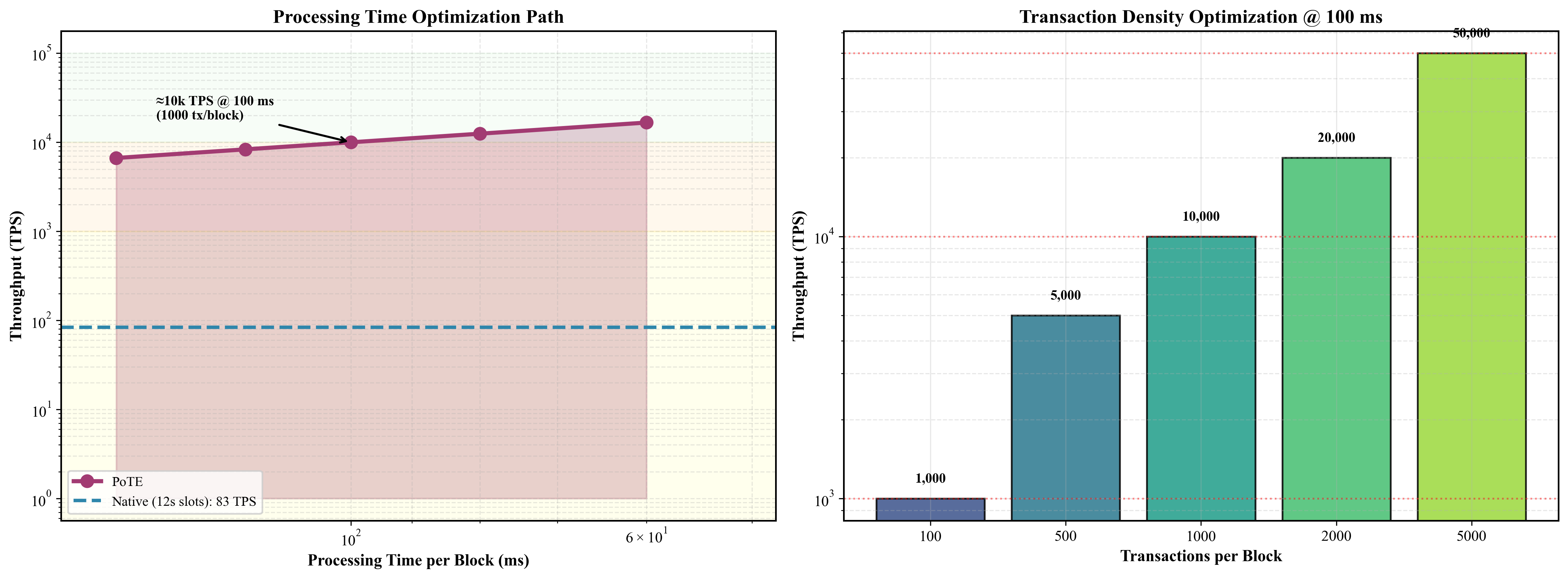}
    \caption{Throughput scalability of PoTE under different optimization targets.} 
	\label{fig:tps}
\end{figure*} 

\subsection{Results Analysis}
Figure \ref{fig:commit-latency} compares the commit latency for the PoTE solution and the native setup as the number of validators increases. The commit latency results  highlight a central
advantage of PoTE: finality no longer depends on slot timing or multi-round
committee voting. In the native consensus configuration, commit latency remains consistently high---between roughly 380\,ms and 450\,ms in our setup---and grows slowly as validators are added. This behavior reflects the structural latency floor imposed by slot scheduling and fork-choice stabilization: even under ideal network conditions, the protocol cannot finalize blocks faster than its fixed timing parameters allow.
PoTE exhibits a markedly different profile. Commit latency begins near
50--60\,ms and increases gradually with network size, remaining below
150\,ms even at 225 validators. This growth is expected: each additional
validator increases the number of attestations that must be disseminated and
verified, and larger networks incur proportionally higher propagation overheads. Yet, unlike the native design, PoTE is not anchored to a consensus epoch or slot interval. Finality occurs as soon as the proposer’s attested block and the multi-vendor quorum of re-attestations have been exchanged, making commit time a function of enclave execution plus network latency, rather than protocol-imposed waiting periods.
Figure~\ref{fig:tps} illustrates PoTE's scalability envelope along two
orthogonal optimization paths: reducing per-block processing time and increasing
transaction density. The left panel shows how throughput improves as the time
required to execute and attest a block decreases. Because PoTE does not rely on
slot times or committee-based voting, the commit latency is determined solely by
enclave execution and network propagation. Even modest optimizations to the
canonical program \(C\) enable substantial throughput gains: at 100\,ms of processing
time per block, PoTE reaches approximately 10{,}000\,TPS with blocks containing
1{,}000 transactions. This region---highlighted in the shaded band---lies entirely outside
the performance envelope of slot-based PoS protocols, whose 12\,s block cadence
caps throughput at roughly 83\,TPS (dashed line). These results underscore the
structural advantage PoTE inherits from decoupling finality from slot timing.

The right panel holds commit latency fixed at 100\,ms and varies the number of
transactions per block. Because PoTE finalizes in a single round of multi-vendor
attestation, throughput grows nearly linearly with transaction density, reaching
5{,}000\,TPS, 10{,}000\,TPS, 20{,}000\,TPS, and up to 50{,}000\,TPS as blocks expand
from 100 to 5{,}000 transactions. This scaling behavior reflects a crucial property
of attested execution: once the proposer has produced a valid enclave-bound block,
verification cost is essentially independent of the number of transactions it
contains. Unlike traditional BFT or PoS consensus, where larger blocks slow down
committee voting or state re-execution, PoTE amortizes computation across a
single deterministic enclave run. Together, these results demonstrate that PoTE
can support high-frequency settlement and dense transaction workloads while
maintaining sub-second deterministic finality, making it suitable for performance-critical
applications such as decentralized exchanges and real-time cross-chain settlement.

\begin{figure}[h]
	\centering
	\includegraphics [scale=0.3] {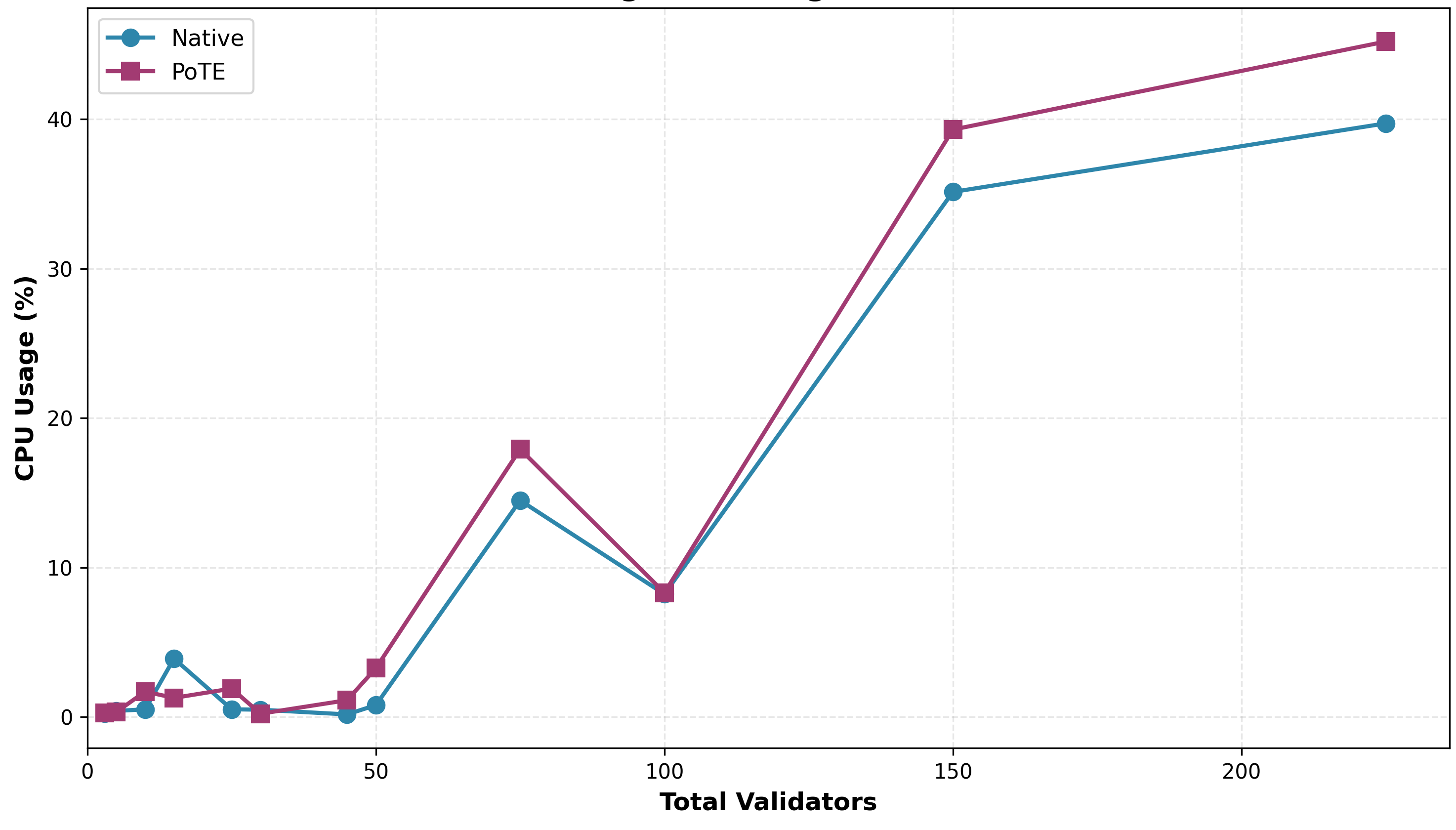}
	\caption{Average CPU Usage}
	\label{fig:cpu}
\end{figure}

The CPU usage results (Figure \ref{fig:cpu}) reveal that PoTE consumes slightly more resources than the native consensus layer. At small validator counts, both systems use very little CPU, but as the network scales past roughly 75–100 validators, CPU load rises sharply for both. The higher CPU usage observed in PoTE is expected and follows directly from the
additional work performed inside each enclave. In the native consensus
configuration, validators primarily execute signature checks, state transitions, and gossip-layer activity. Under PoTE, these tasks are augmented by several CPU-intensive operations: enclaves must verify vendor attestation evidence, bind block contents to the canonical code hash, and generate enclave signatures that anchor the block in hardware. Even when verification is offloaded to the vendor attestation service, validators still perform local quote validation, hashing, and
serialization inside the enclave boundary. These steps are significantly more
expensive than the lightweight cryptographic checks of the native consensus
layer.
\\Moreover, enclave transitions introduce context-switching and memory-isolation overheads not present in standard execution. As the number of validators grows, the network-wide rate of attestations and re-attestations increases, amplifying the aggregate CPU footprint. This effect becomes pronounced beyond approximately 150 validators, where both the enclave's internal cryptographic work and the cost of managing larger state commitments push overall usage upward. 
in the native setting.

\begin{figure}[h]
	\centering
	\includegraphics [scale=0.3] {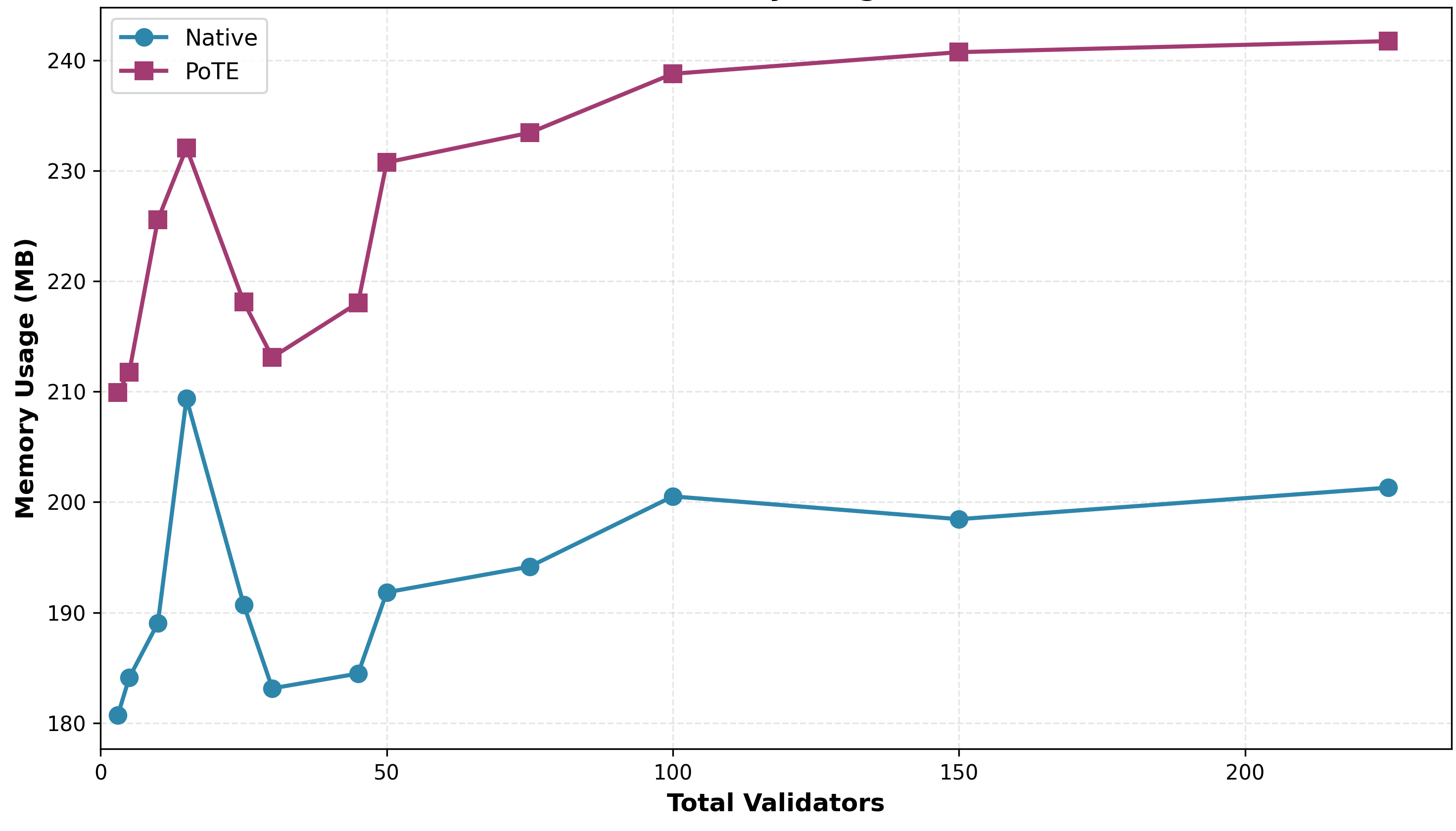}
	\caption{Peak Memory Usage}
	\label{fig:mem}
\end{figure}

For what concerns the memory consumption (Figure \ref{fig:mem}), we can notice that across all validator counts, the PoTE solution uses more memory than the native configuration, with a gap of $~20$-$40$ MB. This overhead reflects the additional buffers, cryptographic structures, and enclave-related state needed when executing inside trusted hardware. Both curves show relatively gentle scaling as the number of validators increases. The native configuration starts at around 180 MB for small validator sets and gradually rises to just over 200 MB by the time the system reaches 225 validators. The TEE configuration follows the same overall trend but begins higher, at roughly 210 MB, and climbs to approximately 242 MB. In the lower part of the range, particularly between 5 and 50 validators, both lines exhibit small oscillations. These fluctuations are typical in small clusters where block production, caching, and gossip overhead can vary more from epoch to epoch. As the validator set grows beyond roughly 100 nodes, both lines flatten and become stable: native execution settles around 198–202 MB, while the TEE mode remains steady near 238–242 MB.

\section{Conclusion}
\label{conclusion}
This paper introduced \emph{Proof of Trusted Execution} (PoTE), a consensus
architecture built around deterministic, attested execution rather than
probabilistic agreement or economically weighted voting. By executing a canonical
program \(C\) inside heterogeneous TEEs and requiring corroborating attestations
from multiple independent vendors, PoTE replaces the structural latency
constraints of Proof-of-Stake and Proof-of-Work with a single-round, hardware
anchored finality mechanism. The resulting protocol admits no forks, produces a
unique block at each height, and commits state transitions at latencies dominated
only by attestation generation and network propagation.
\\Our implementation demonstrates that a multi-vendor federation of TEEs can
achieve sub-second deterministic settlement while preserving auditability and
without altering the execution semantics visible to applications. This makes PoTE
a natural fit for high-frequency or latency-bound decentralized systems such as
matching engines, cross-chain settlement layers, or autonomous financial logic
that cannot tolerate the slot-time ceilings, multi-round voting, or reorg risks of
conventional PoS designs.
\\TEEs, however, are not infallible. Attacks such as \emph{tee.fail}, Foreshadow,
Plundervolt, and other physical or microarchitectural exploits have shown that
specific implementations can be compromised under determined adversaries. PoTE
addresses this not by assuming perfect hardware, but by distributing trust across
independent vendors and allowing the system to incorporate future TEE
technologies as they mature. The design is flexible enough to accommodate VM-
based platforms such as Intel TDX, AMD SEV, AWS Nitro, or upcoming ARM CCA
environments, enabling continual evolution of the trust base as the ecosystem
advances.
\\Taken together, our results suggest that hardware-rooted, deterministic execution
offers a viable alternative to traditional consensus paradigms and opens a
promising direction for building low-latency, high-assurance, and
cryptographically accountable distributed systems. Future work includes
formalizing PoTE’s security properties under richer threat models, exploring
vendor-diverse enclave committees, and integrating lightweight zero-knowledge
proofs to further strengthen transparency and resilience.

\begin{acks}
\end{acks}

\bibliographystyle{ACM-Reference-Format}
\bibliography{references}

\end{document}